\newcommand{\dedx}{dE/dx}

\newcommand{\llb}{\Lambda\bar{\Lambda}}

\newcommand{\lm}{\Lambda}
\newcommand{\lmb}{\bar{\Lambda}}

\newcommand{\GG}{\gamma\gamma}

\newcommand{\pppr}{\pi^+\pi^-p\bar{p}}

\newcommand{\jpsi}{J/\psi}
\newcommand{\ar}{\rightarrow}

\newcommand{\bfg}{\begin{figure}}
\newcommand{\efg}{\end{figure}}
\newcommand{\bitm}{\begin{itemize}}
\newcommand{\eitm}{\end{itemize}}
\newcommand{\bnum}{\begin{enumerate}}
\newcommand{\enum}{\end{enumerate}}
\newcommand{\btbl}{\begin{table}}
\newcommand{\etbl}{\end{table}}
\newcommand{\btbu}{\begin{tabular}}
\newcommand{\etbu}{\end{tabular}}
\newcommand{\bcl}{\begin{center}}
\newcommand{\ecl}{\end{center}}
\newcommand{\bbt}{\bibitem}
\newcommand{\beq}{\begin{equation}}
\newcommand{\eeq}{\end{equation}}
\newcommand{\beqr}{\begin{eqnarray}}
\newcommand{\eeqr}{\end{eqnarray}}
\documentclass[preprint,showpacs,amsmath,amssymb]{revtex4}
\usepackage{rotating,epsfig,graphicx}
\usepackage{flafter}
\include{def-com}
\normalsize
\begin{document}
\title{\boldmath \bf First measurements of $\jpsi$ decays into $\Sigma^+ \bar{\Sigma}^-$ and $\Xi^0 \bar{\Xi}^0$ }
\author{
M.~Ablikim$^{1}$,    J.~Z.~Bai$^{1}$,    Y.~Bai$^{1}$,
Y.~Ban$^{11}$,   X.~Cai$^{1}$,   H.~F.~Chen$^{16}$,
H.~S.~Chen$^{1}$,   H.~X.~Chen$^{1}$,   J.~C.~Chen$^{1}$,
Jin~Chen$^{1}$
% ...
\\
\vspace{0.2cm}
(BES Collaboration)\\
\vspace{0.2cm}
{\it
$^{1}$ Institute of High Energy Physics, Beijing 100049, People's Republic of China\\
% ...
$^{11}$ Peking University, Beijing 100871, People's Republic of China\\
% ...
$^{16}$ University of Science and Technology of China, Hefei 230026,
People's Republic of China\\
% ...
}
}
\date{\today}

\begin{abstract}
Based on 58 million $J/\psi$ events collected with the BESII
detector at the BEPC, the baryon pair processes $J/\psi\ar\Sigma^+
\bar{\Sigma}^-$ and $J/\psi\ar\Xi^0 \bar{\Xi}^0$ are observed for
the first time. The branching fractions are measured to be ${\cal
B}(\jpsi\ar\Sigma^+ \bar{\Sigma}^-)=(1.50\pm 0.10\pm 0.22)\times
10^{-3}$ and ${\cal B}(\jpsi\ar\Xi^0 \bar{\Xi}^0)=(1.20\pm 0.12\pm
0.21)\times 10^{-3}$, where the first errors are
statistical and the second ones are systematic.
\end{abstract}
\pacs{13.25.Gv, 12.38.Qk, 14.20.Gk, 14.40.Cs}
\maketitle
\section{Introduction}
Since the discovery of the charmonium states $J/\psi$ and $\psi(2S)$,
a number of baryonic decay channels have been studied by
several different
experiments~\cite{np1,np2,np3,np4,np5,np6,np7,np8,np9,np10,np11,np12}.
Baryon-antibaryon decays, which provide a
 test of the predictive
power of QCD, have especially attracted
 interest of both theoretical
and experimental experts. Among these
 decays, $J/\psi\ar\Sigma^0
\bar{\Sigma}^0$ and $J/\psi\ar\Xi^-
 \bar{\Xi}^+$ have been studied
by DM2 and
 MarkII~\cite{np2,np3}, but their isospin partners decays
$J/\psi\ar\Sigma^+ \bar{\Sigma}^-$ and $J/\psi\ar\Xi^0
 \bar{\Xi}^0$
have not been measured before. In this article, we study
$J/\psi\ar\Sigma^+ \bar{\Sigma}^-$ and
 $J/\psi\ar\Xi^0 \bar{\Xi}^0$
using the large $J/\psi$ data sample accumulated with the BESII
detector. The
 decay mode $J/\psi\ar\Sigma^- \bar{\Sigma}^+$ is not
studied here since the final states contain a neutron and an
antineutron, which are difficult to detect with the BESII
detector. Isospin invariance predicts ${\cal B}(\Sigma^+
\bar{\Sigma}^-)={\cal B}(\Sigma^0 \bar{\Sigma}^0)$ and ${\cal B}(\Xi^0
\bar{\Xi}^0)={\cal B}(\Xi^- \bar{\Xi}^+)$. However in the quark model,
there are well-known isospin breaking contributions in $J/\psi$
baryonic decays \cite{np21,np22}.

\section{The BESII Detector and Monte Carlo simulation}
BESII is a conventional solenoidal magnet detector that is
described in detail in Ref.~\cite{np13}. A 12-layer vertex chamber
(VTC) surrounding the beam pipe provides trigger and track information. A
forty-layer main drift chamber (MDC), located radially outside the
VTC, provides trajectory and energy loss ($dE/dx$) information for
charged tracks over $85\%$ of the total solid angle.  The momentum
resolution is $\sigma _p/p = 0.017 \sqrt{1+p^2}$ ($p$ in
$\hbox{GeV}/c$), and the $dE/dx$ resolution for hadron tracks is
$\sim 8\%$. An array of 48 scintillation counters surrounding the
MDC measures the time-of-flight (TOF) of charged tracks with a
resolution of $\sim 200$ ps for hadrons.  Radially outside the TOF
system is a 12 radiation length, lead-gas barrel shower counter
(BSC). This measures the energies of electrons and photons over
$\sim 80\%$ of the total solid angle with an energy resolution of
$\sigma_E/E=22\%/\sqrt{E}$ ($E$ in GeV). Outside of the solenoidal
coil, which provides a 0.4~Tesla magnetic field over the tracking
volume, is an iron flux return that is instrumented with three
double layers of counters that identify muons of momentum greater
than 0.5~GeV/$c$.

In this analysis, a GEANT3 based Monte Carlo (MC) simulation program~\cite{np14} with detailed consideration of real detector responses (such as dead electronic channels) is used. The consistency between data and MC simulation has been carefully checked in many high-purity physics channels, and the agreement is quite reasonable~\cite{np15}.
\section{Event selection}
The data sample used for this analysis consists of $(58.0\pm
2.7)\times 10^6$ $\jpsi$ events collected with the BESII detector~\cite{np23}.
The decay channels investigated are $J/\psi$ decays into $\Sigma^+
\bar{\Sigma}^-$ and $\Xi^0 \bar{\Xi}^0$ baryon pairs, where
$\Sigma^+$ decays to $\pi^0 p$ ($\pi^0\ar\GG$), $\Xi^0$ to
$\pi^0\Lambda$ ($\Lambda\ar\pi^-p$). Therefore the final states for
the two decays are $p\bar{p}\GG\GG$ and $\pi^+ \pi^-
p\bar{p}\GG\GG$, respectively. Both decays contain four photons
 in
the final states. Candidate events are required to satisfy the
following common
 selection criteria:
\begin{enumerate}
\item Events must have two or four good charged tracks with zero net
charge. A good charged track is a track that is well fitted to a helix
in the MDC, and has a polar angle, $\theta$, in the range $|\cos
\theta|<0.8$. For $J/\psi\ar\Sigma^+\bar{\Sigma}^-$, tracks are
required to originate from the interaction region of $R_{xy} <0.02$ m
and $|z| < 0.2$ m, where $R_{xy}$ is the distance from the beamline to
the point of closest approach of the track to the beamline, and $|z|$
is the distance along the beamline to this point from the interaction
point. For $J/\psi\ar\Xi^0\bar{\Xi}^0$, because of the long lifetime
of $\Xi$ and $\Lambda$, tracks are not required to originate from the
interaction region.
\item The TOF and $\dedx$ measurements of the charged tracks are used
to calculate $\chi^{2}_{PID}$ values for the hypotheses that the
particle is a pion, kaon, or proton. Only the two proton tracks must
be identified with the requirement that $\chi^{2}_{PID}$ for the
proton hypothesis is less than those for the $\pi$ or $K$
hypotheses.
\item Isolated photons are those that have energy deposited in the BSC
greater than 50 MeV, and the angle between the photon entering
 the
BSC and the shower development direction in the BSC is less than
$37^{\circ}$. In order to remove the fake photons produced by
$\bar{p}$
 annihilation and those produced by hadronic interactions
of tracks, the angle between the photon and
 antiproton is required
to be larger than $25^{\circ}$ and those between the photon and other
charged tracks larger than $8^{\circ}$.
\end{enumerate}
\begin{figure*}
\centerline{\psfig{file=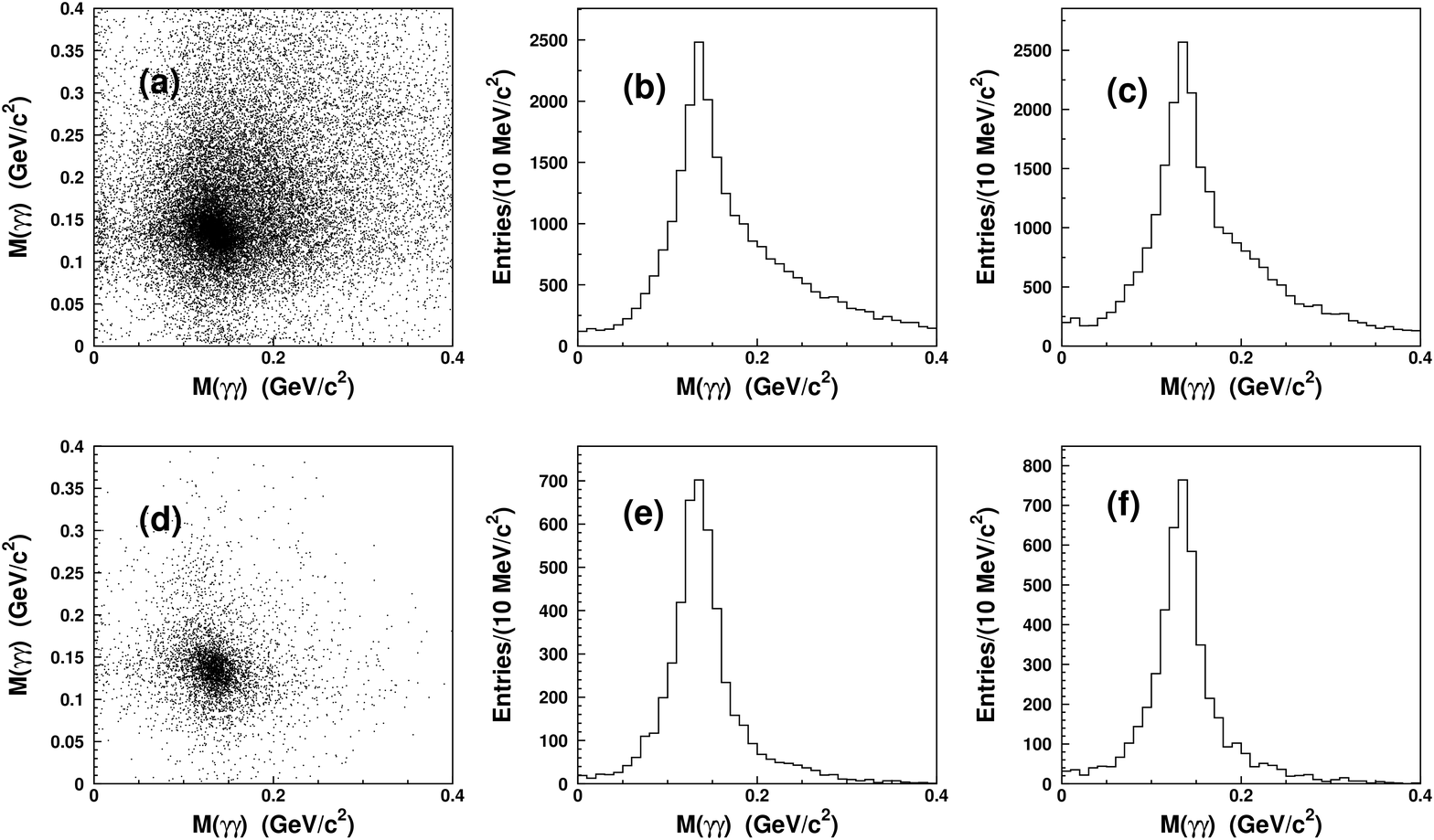,width=16cm,height=12cm}}
\caption{Invariant mass distributions of two photons for
  $\jpsi\ar\Sigma^+\bar{\Sigma}^-$ candidate events. (a) Scatter plot
of $M(\GG)_1$ versus $M(\GG)_2$ for the combination with the minimum
$R(\pi^0)$, (b) distribution of $M(\GG)_1$, (c) distribution of
$M(\GG)_2$ in the data sample. and (d), (e), and (f) are the
corresponding plots for MC simulated $\jpsi\ar\Sigma^+\bar{\Sigma}^-$
events. }
\label{fig1}
\end{figure*}

Different kinematic fits are used in the selection of the two decay
channels. A four constraint (4C) kinematic fit under
the $p\bar{p}\GG\GG$ hypothesis is performed for $J/\psi\ar\Sigma^+
\bar{\Sigma}^-$. If there are more than four photon candidates in an
event, all combinations are tried, and the combination with the
smallest $\chi^2_{4C}$ is retained. We require the minimum
$\chi^2_{4C}$ to be less than 15.  For $\jpsi\ar\Xi^0\bar{\Xi}^0$, a
six constraint (6C)
 kinematic fit under the hypothesis
$\jpsi\ar\GG\GG\pi^+\pi^- p\bar{p}$ with the
 invariant mass of the
two photon pairs constrained to the $\pi^0$
 mass is performed, and
the $\chi^2$
 of the 6C fit is required to be less than 50.
\section{Data Analysis}
\subsection{$\jpsi\ar\Sigma^+\bar{\Sigma}^-$}
The candidate events for this decay mode contain two $\pi^0$, and there
are three possible combinations of $(\gamma \gamma)_1$
 $(\gamma
\gamma)_2$ to form a $\pi^0$ pair. The $\pi^0$ pair
 with the
minimum $R(\pi^0)$, where $R(\pi^0)=\sqrt{(M(\gamma
\gamma)_1-M(\pi^0))^2+(M(\gamma \gamma)_2-M(\pi^0))^2}$,
 is chosen
for further analysis. Figure~\ref{fig1} shows the mass distributions
of candidate events with the minimum $R(\pi^0)$ for data and MC
samples, respectively. A clear $\pi^0 \pi^0$ signal is observed in the
data sample. In order to select
 $\pi^0$ pair events,
$|M(\GG)_1-M(\pi^0)|<0.03$ GeV/$c^2$ and
 $|M(\GG)_2-M(\pi^0)|<0.03$
GeV/$c^2$ are required.

\begin{figure*}
\centerline{\psfig{file=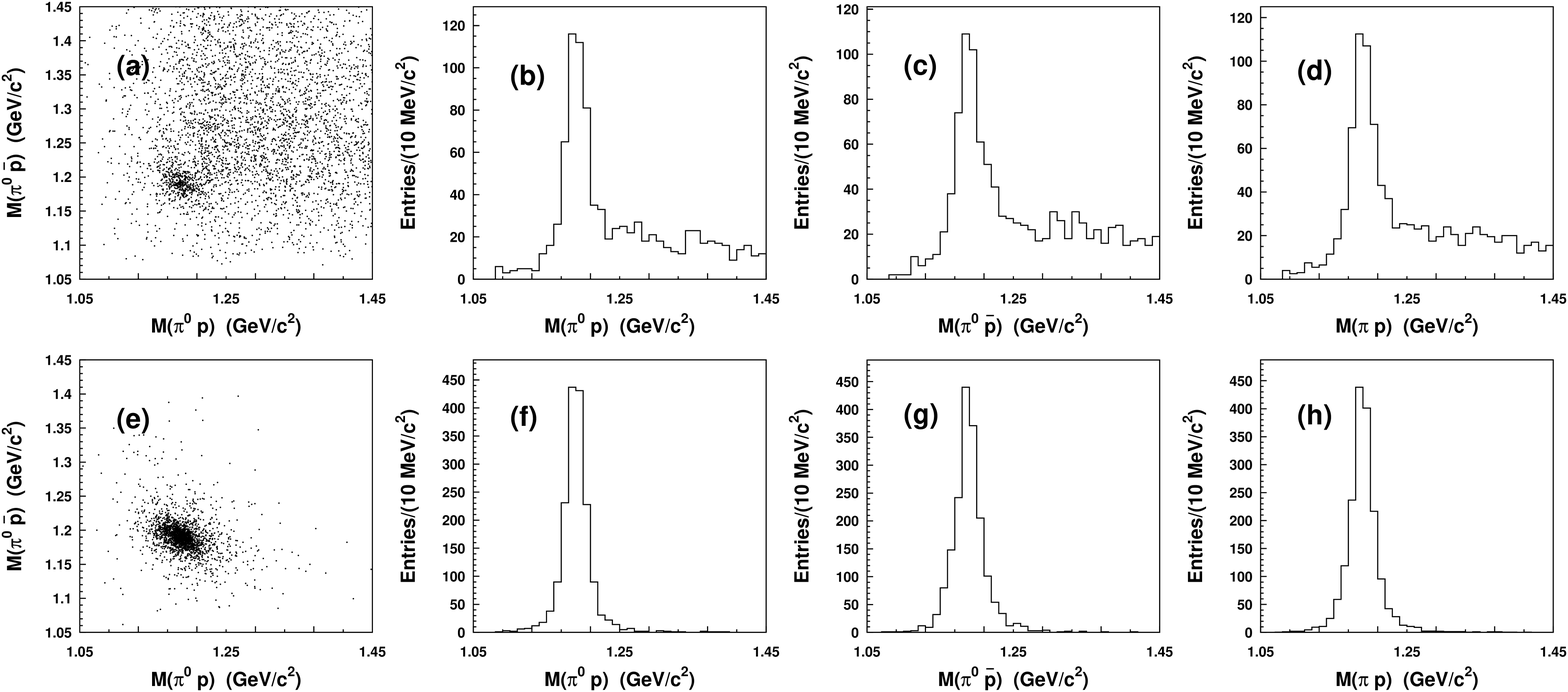,width=16cm,height=8cm}}
\caption{Distributions for $\jpsi\ar\Sigma^+\bar{\Sigma}^-$
candidate
  events. (a) Scatter plot of $M(\pi^0_{(1)} p)$ versus $M(\pi^0_{(2)}
  \bar{p})$; (b) $M(\pi^0_{(1)} p)$ distribution by requiring
  $|M(\pi^0_{(2)} \bar{p})-M(\Sigma^+)|<0.03$ GeV/$c^2$, (c)
  $M(\pi^0_{(2)} \bar{p})$ distribution by requiring $|M(\pi^0_{(1)}
  p)-M(\Sigma^+)|<0.03$ GeV/$c^2$, and (d) is the sum of (b) and (c)
  scaled by a factor of 0.5. (e), (f), (g) and (h) are the
  corresponding plots of (a), (b), (c) and (d) for MC simulated
  $\jpsi\ar\Sigma^+\bar{\Sigma}^-$ events.}
\label{fig2}
\end{figure*}

After $\pi^0$ selection, there are still two possible $\pi^0 p$
combinations from which to form the $\Sigma$ of the $\Sigma^+
\bar{\Sigma}^-$ pair. The combination
 having the smallest value of
$R(\Sigma)=\sqrt{(M(\pi^0_{(1)} p)-M(\Sigma))^2+(M(\pi^0_{(2)}
\bar{p})-M(\Sigma))^2}$ is selected for further
analysis. Figure~\ref{fig2} shows the $\pi p$ invariant mass for data
and MC simulated $\jpsi\ar\Sigma^+
 \bar{\Sigma}^- $ events. A clear
$\Sigma^+ \bar{\Sigma}^-$ signal is seen in the bottom left corner of
Fig.~\ref{fig2}(a). Figure~\ref{fig2}(b) is the
 $M(\pi^0_{(1)}p)$
distribution by requiring
 $|M(\pi^0_{(2)}\bar{p})-M(\Sigma)|<0.03$
GeV/$c^2$, (c) is the
 $M(\pi^0_{(2)}\bar{p})$ distribution by
requiring $|M(\pi^0_{(1)}
 p)-M(\Sigma)|<0.03$ GeV/$c^2$, and (d) is
the sum of (b) and (c) scaled by a factor of 0.5.

Possible backgrounds come
 from channels with $p\bar{p}$ production,
including
 $\jpsi\ar p\bar{p}$, $\jpsi\ar\gamma p\bar{p}$,
$\jpsi\ar
 p\bar{p}\pi^0$, $\jpsi\ar p\bar{p}\eta$ ($\eta\ar\GG$ or
$\eta\ar
 3\pi^0$), $\jpsi\ar p\bar{p}\omega$
($\omega\ar\gamma\pi^0$), and
 $\jpsi\ar p\bar{p}\pi^0 \pi^0$. MC
events are generated with a phase space generator for the first two
background channels.  None or only a few events survive the selection
criteria; therefore contamination from the first two channels is
negligible. For $\jpsi\ar p\bar{p}\pi^0$, $\jpsi\ar p\bar{p}\eta$, and
$\jpsi\ar p\bar{p}\omega$, MC events are
 also generated according to
phase space. Using the branching
 fractions from the PDG~\cite{np16},
the numbers of events from these channels
 are expected to be 7.2,
72.4 and 7.8 in the whole $\pi p$ mass region, respectively. For
$\jpsi\ar
 p\bar{p}\pi^0 \pi^0$, the branching fraction is
unavailable, so the normalized number of events for this background
can not be determined. However, the $\pi p$ invariant mass
distribution from all backgrounds is smooth, so these backgrounds
will not affect the determination of the number of signal events in
fitting the $\pi p$ mass distribution.  \bfg
\centerline{\psfig{file=fig3.epsi,width=8cm,height=8cm}}
\caption{Fit to $\pi p$ invariant mass of
  $\jpsi\ar\Sigma^+\bar{\Sigma}^-$ candidate events with MC simulated
  signal shape and a second order polynomial as background shape. The
  shaded histogram is background from MC simulated $\jpsi\ar
  p\bar{p}\pi^0$, $\jpsi\ar p\bar{p}\eta$, and $\jpsi\ar
  p\bar{p}\omega$, normalized according to the branching fractions in
  PDG~\cite{np16}. The dashed histogram shows the shape of MC
  simulated $J/\psi\ar p\bar{p}\pi^0\pi^0$ normalized using an
  assigned branching fraction of 0.003.}
\label{fig3}
\efg

In order to determine the branching fraction, we fit the $\Sigma$ signal in
Fig.~\ref{fig2}(d) with a histogram of the signal shape from MC
simulation together with a second order polynomial for the
background. The fit is shown in Fig.~\ref{fig3}, and it
yields $399 \pm 26$ signal events, with the goodness of the fit being
$\chi^2/ndf=22.4/31\approx 0.72$. The $J/\psi\ar\Sigma^+
\bar{\Sigma}^-$ efficiency is determined to be $\varepsilon$=1.75\% using MC
simulated signal events, and the branching fraction is,
\begin{equation*}
\begin{array}{ll}
{\cal B}(\jpsi\ar\Sigma^+\bar{\Sigma}^-)&={\displaystyle \frac{N(\Sigma)/ \varepsilon}
{N(\jpsi)\cdot{\cal B}^2(\Sigma^+\ar\pi^0 p)\cdot{\cal B}^2(\pi^0\ar\GG)}}\\
&=(1.50 \pm 0.10)\times 10^{-3},
\end{array}
\end{equation*}
where the error is statistical.

\subsection{$\jpsi\ar\Xi^0\bar{\Xi}^0$}
The candidate events for this decay mode contain a
$\Lambda\bar{\Lambda}$ pair. In order to select $\Lambda\bar{\Lambda}$
events, we require the $\pi^- p$ and $\pi^+ \bar{p}$ invariant masses
satisfy $|M(\pi^- p)-M(\Lambda)|<0.01$GeV/$c^2$ and $|M(\pi^+
\bar{p})-M(\Lambda)|<0.01$GeV/$c^2$.

\begin{figure*}
\centerline{\psfig{file=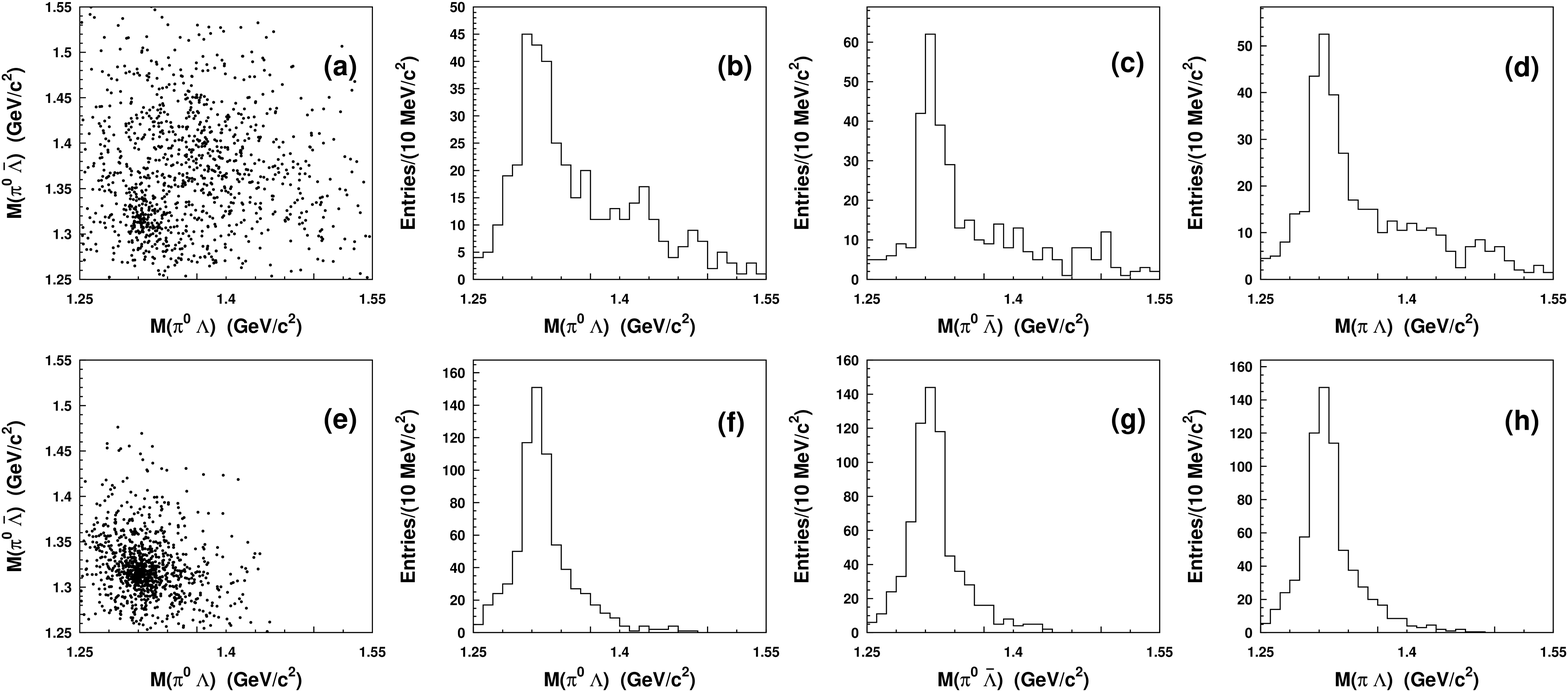,width=16cm,height=8cm}}
\caption{Plots for $\jpsi\ar\Xi^0\bar{\Xi}^0$ candidate events.  (a)
is the scatter plot of $M(\pi^0_{(1)} \lm)$ versus $M(\pi^0_{(2)}
\lmb)$, (b) is the $M(\pi^0_{(1)}\lm)$ distribution recoiling
against $\bar{\Xi}^0$, selected by requiring
$|M(\pi^0_{(2)}\lmb)-M(\Xi^0)|<0.03$GeV/$c^2$, (c) is the
$M(\pi^0_{(2)}\lmb)$ distribution recoiling against $\Xi^0$,
selected by requiring $|M(\pi^0_{(1)}\lm)-M(\Xi^0)|<0.03$GeV/$c^2$,
(d) is the sum of (b) and (c) scaled by a factor of $0.5$, (e), (f),
(g) and (h) are the corresponding plots of (a), (b), (c), and (d)
for MC simulated $\jpsi\ar\Xi^0\bar{\Xi}^0$ events.} \label{fig4}
\end{figure*}

There are three possible combinations of the four photons to form a $\pi^0$
pair in the 6C kinematic fit; the combination with the minimum
$\chi^2_{6C}$ is considered to be the correct one and selected for
further investigation. After $\pi^0$ pairs are selected, there are
two possible combinations ($\pi^0_{(1)}\Lambda$, $\pi^0_{(2)}
\Lambda$) to form $\Xi^0$ candidates. Analogous to the analysis of
$\Sigma^+ \bar{\Sigma}^-$, we choose the combination with the lowest
value of
$R(\Xi^0)=\sqrt{(M(\pi^0_{(1)}\Lambda)-M(\Xi^0))^2+(M(\pi^0_{(2)}\bar{\Lambda})-M(\Xi^0))^2}$
for further study. Figure~\ref{fig4} shows the $\pi\Lambda$ invariant mass
for this case for data and MC simulated
$\jpsi\ar\Xi^0\bar{\Xi}^0$ events. In Fig.~\ref{fig4} (a), besides the
clear $\Xi^0 \bar{\Xi}^0$, $\Sigma(1385)^0 \bar{\Sigma}(1385)^0$
production is also visible. In Fig.~\ref{fig4} (b) is the
$M(\pi^0_{(1)}\Lambda)$ distribution after requiring $|M(\pi^0_{(2)}
\bar{\Lambda})-M(\Xi^0)|<0.03$ GeV/$c^2$, (c) is the
$M(\pi^0_{(2)}\bar{\Lambda})$ distribution requiring
$|M(\pi^0_{(1)} \Lambda)-M(\Xi^0)|<0.03$ GeV/$c^2$, and (d) is the sum
of (b) and (c) scaled by a factor of $0.5$.

Possible backgrounds come from channels with $\Lambda$ or $\Xi$
production, including $\jpsi\ar\Sigma^0 \bar{\Sigma}^0$,
$\jpsi\ar\Sigma^0\pi^0\bar{\Lambda}(+c.c.)$, $\jpsi\ar\Sigma(1385)^0
\bar{\Sigma}(1385)^0$, and $\jpsi\ar\pi^0\pi^0\Lambda\bar{\Lambda}$.
Using the branching fraction of $\jpsi\ar\Sigma^0 \bar{\Sigma}^0$
from the PDG~\cite{np16} and assuming isospin invariance holds for
$\jpsi\ar\Sigma\pi\bar{\Lambda}+c.c.$ and $\jpsi\ar\Sigma(1385)
\bar{\Sigma}(1385)$, we obtain 0.6, 45.9, and 51.3 background events
from MC simulation, respectively. The shaded part in Fig.~\ref{fig5}
shows the normalized simulated backgrounds from the first three
channels, which do not exhibit any peaking structures in the $\Xi^0$
mass region. Since the branching fraction is unavailable for
$\jpsi\ar\pi^0\pi^0\Lambda\bar{\Lambda}$, we do not determine the
normalized number of events for this decay mode. However, MC
simulation indicates that the $\pi\Lambda$ invariant mass
distribution is smooth without any peaking, and therefore this
background will not affect the determination of the number of signal
events. We also studied backgrounds from $\llb$ sidebands and other
possible background channels listed in the PDG, but the
contaminations were found to be negligible. The fitted number of
signal events is insensitive to the shape of all the backgrounds
considered here. The numbers of signal events using different
background shapes differ slightly from each other; we consider this
difference as one source of systematic error.  \bfg
\centerline{\psfig{file=fig5.epsi,width=8cm,height=8cm}}
\caption{Fit to $\pi\lm$ invariant mass spectrum of
  $\jpsi\ar\Xi^0\bar{\Xi}^0$ candidate events. Dots with error bars
  are data, the hatched histogram is the normalized background from
  all the channels considered in the text, and the solid histogram is
  the fit to data using a histogram of the signal shape from MC
  simulation plus a second order polynomial for background.}
\label{fig5}
\efg

To get the number of signal events, we fit the observed $\Xi$ signal
in Fig.~\ref{fig4} (d) by a histogram of the signal shape from MC
simulation plus a second order polynomial as the background. The
fitting result is shown in Fig.~\ref{fig5} and the fit yields $206
\pm20$. The efficiency is determined to be $\varepsilon$=0.74\% using
MC simulated signal events generated according to a phase space
distribution. The branching fraction is,
\begin{equation*}
\begin{array}{ll}
{\cal B}(\jpsi\ar\Xi^0\bar{\Xi}^0)&={\displaystyle \frac{N(\Xi)/ \varepsilon}
{N(\jpsi)\cdot{\cal B}^2(\Xi^0\ar\pi^0\lm)\cdot
{\cal B}^2(\lm\ar\pi^-p)\cdot{\cal B}^2(\pi^0\ar\GG)}}\\
&=(1.20 \pm 0.12)\times 10^{-3},
\end{array}
\end{equation*}
where the error is statistical.
\begin{table*}
\caption{Summary of systematic errors (\%).}
\bcl
\begin{tabular}{l|c|c}\hline\hline
Source&$\jpsi\ar\Sigma^+\bar{\Sigma}^-$&$\jpsi\ar\Xi^0\bar{\Xi}^0$ \\ \hline
MDC tracking and PID&2&6 \\
Photon efficiency&8&8\\
Kinematic fit&4&8.4  \\
Background shape&6&5 \\
$\jpsi$ statistics&4.7&4.7  \\
$\alpha$&8&7 \\
VC trigger efficiency&1&4   \\ \hline
Total&14.4&16.9 \\ \hline
\end{tabular}
\label{table-1}
\ecl
\end{table*}
\section{Systematic errors}
The systematic errors on the branching ratios mainly arise from the
uncertainties in the MDC tracking, particle identification, photon
efficiency, angular distribution parameter $\alpha$ in event generators, kinematic fitting, background
shapes, and the total number of $\jpsi$ events. The errors from different
sources are listed in Table~\ref{table-1}.

The uncertainties caused by MDC tracking and particle identification (PID) are estimated
by the difference of the selection efficiency of proton and antiproton
between data and MC simulation~\cite{np17}. The efficiencies of PID
and track reconstruction for protons and antiprotons that enter the
detector being reconstructed and identified are measured using samples
of $\jpsi\ar\pppr$, which are selected using PID for three tracks,
allowing one proton or antiproton at a time to be missing in the
fit~\cite{np17}. It is found that the efficiency difference of one
proton identification is about 1\% for
$\jpsi\ar\Sigma^+\bar{\Sigma}^-$ and about 2\% for
$\jpsi\ar\Xi^0\bar{\Xi}^0$ depending on the momentum of the final
state particles. The $\pi^{\pm}$ tracking and PID efficiencies are simulated within 1\% per track. Therefore we get a total of 2\% for
$\jpsi\ar\Sigma^+\bar{\Sigma}^-$ and 6\% for
$\jpsi\ar\Xi^0\bar{\Xi}^0$, respectively.

The photon detection efficiency is studied using $\jpsi\ar\rho^0\pi^0$ in
Ref.~\cite{np18}. The results indicate that the systematic error is
about 2\% for each photon. Therefore, 8\% is taken as the systematic
error of photon efficiency for the two decay modes.

The angular distribution of the baryon in $J/\psi$ decay is $1+\alpha \cos^2
\theta$, with $\theta$ being the polar angle of the baryon in $J/\psi$
rest frame. To estimate the uncertainty originating from the angular distribution
parameter $\alpha$, we generate MC samples for $\alpha=0$ and
$\alpha=1$, separately. The differences of efficiency between
$\alpha=0$ and $\alpha=1$ are taken as systematic errors, which is 8\%
for $\jpsi\ar\Sigma^+\bar{\Sigma}^-$ and 7\% for
$\jpsi\ar\Xi^0\bar{\Xi}^0$.

The systematic errors for the kinematic fits are 4\% for
$\Sigma^+\bar{\Sigma}^-$ and 8.4\% for $\Xi^0\bar{\Xi}^0$, they are
taken from earlier studies~\cite{np19,np20}. The VTC trigger efficiency
systematic errors are estimated to be 1\% for
$\Sigma^+\bar{\Sigma}^-$ and 4\% for $\Xi^0\bar{\Xi}^0$.

The systematic error of the background shape used is estimated by measuring
the difference of the numbers of fitted signal events for different
background shapes. For $\jpsi\ar\Sigma^+\bar{\Sigma}^-$, we also fit
the $\pi p$ invariant mass distribution using the normalized
background, or MC simulated $\jpsi\ar\pi^0\pi^0p\bar{p}$ events as the
background shape. The difference of the numbers of signal events is
about 6\% compared to our nominal fit, which is taken as the
systematic error of the background shape. For
$\jpsi\ar\Xi^0\bar{\Xi}^0$, we fit the $\pi \Lambda$ invariant mass
distribution using the normalized background or MC simulated
$\jpsi\ar\pi^0\pi^0\Lambda\bar{\Lambda}$ events as the background
shape, and we estimate a systematic error of about 5\%.

Uncertainty on the total number of $\jpsi$ events is 4.7\%~\cite{np23}.
Combining these errors in quadrature gives total systematic errors of
14.4\% for $\jpsi\ar\Sigma^+\bar{\Sigma}^-$ and 16.9\% for $\jpsi\ar\Xi^0\bar{\Xi}^0$.

\section{Results and discussion}
Based on $58\times 10^6 \jpsi$ events accumulated at BESII, we report
first measurements of the branching fractions of $\jpsi$ decays into
the baryon pairs $\Sigma^+\bar{\Sigma}^-$ and $\Xi^0\bar{\Xi}^0$. The
results are listed in Table~\ref{table-2}, including the results of
$\jpsi\ar\Sigma^0\bar{\Sigma}^0$ and $\jpsi\ar\Xi^-\bar{\Xi}^+$. We
note that the isospin partners, $\Sigma^+$ and $\Sigma^0$ and also
$\Xi^0$ and $\Xi^-$, have similar branching fractions in agreement
with expectations of isospin symmetry.  Furthermore, according to the
phase space corrected branching fraction $|M_i|^2={\cal B}(\jpsi\ar
B_i\bar{B}_i)/(\pi p^*/\sqrt{s})$~\cite{np5}, we obtain the phase
space corrected branching fractions $(1.50\pm0.10\pm0.22)$ and
$(1.45\pm0.15\pm0.26)$ for $\jpsi\ar\Sigma^+\bar{\Sigma}^-$ and
$\jpsi\ar\Xi^0\bar{\Xi}^0$, respectively.  We note
that the increase of strangeness does not greatly change the branching
fraction, indicating the flavor symmetric nature of gluons. We also
calculate the ratios of the $\psi(2S)$ results in Ref.~\cite{np6} to
those from $J/\psi$ measurements in this article after removing the phase space factor,
and obtain $(14.3\pm 5.2)$\% for $\Sigma^+\bar{\Sigma}^-$ and $(17.3\pm 6.7)$\% for
$\Xi^0\bar{\Xi}^0$. They agree with the so called "12\% rule"
predicted by perturbative QCD~\cite{np24} within $1\sigma$.
\begin{table*}
\caption{Branching fractions ($10^{-3}$) of $\jpsi$ decays into
$\Sigma^+\bar{\Sigma}^-$, $\Sigma^0\bar{\Sigma}^0$,
$\Xi^0\bar{\Xi}^0$, and
$\Xi^-\bar{\Xi}^+$~\cite{np1,np2,np3,np7,np11}. The first error is
statistical and the second systematic. For the results of
MarkI~\cite{np1}, the statistical and systematic errors were added
in quadrature.}  \bcl
\begin{tabular}{l|c|c|c|c}\hline\hline
Channels&$\jpsi\ar\Sigma^0\bar{\Sigma}^0$&$\jpsi\ar\Xi^-\bar{\Xi}^+$&$\jpsi\ar\Sigma^+\bar{\Sigma}^-$&$\jpsi\ar\Xi^0\bar{\Xi}^0$
\\ \hline
MarkI&$1.3\pm0.4$&$1.4\pm0.5$& &$3.2\pm0.8$($\Xi^0\bar{\Xi}^0+\Xi^-\bar{\Xi}^+$) \\
MarkII&$1.58\pm0.16\pm0.25$&$1.14\pm0.08\pm0.20$& & \\
DM2&$1.06\pm0.04\pm0.23$&$1.4\pm0.12\pm0.24$& &  \\
BESII&$1.33\pm0.04\pm0.11$& &$1.50\pm0.10\pm0.22$&$1.20\pm0.12\pm0.21$ \\
BaBar&$1.15\pm0.24\pm0.03$& & &  \\ \hline
\end{tabular}
\label{table-2}
\ecl
\end{table*}
\section{Acknowledgments}
The BES collaboration thanks the staff of BEPC and computing center
for their hard efforts. This work is supported in part by the
National Natural Science Foundation of China under contracts Nos.
10491300, 10825524, 10225524, 10225525, 10425523,10625524, 10521003,
the Chinese Academy of Sciences under contract No. KJ 95T-03, the
100 Talents Program of CAS under Contract Nos. U-11, U-24, U-25, and
the Knowledge Innovation Project of CAS under Contract Nos. U-602,
U-34 (IHEP), U-612(IHEP), the National Natural Science Foundation of
China under Contract Nos. 10225522, 10491305 (Tsinghua University),
MOE of China under contract No. IRT0624 (CCNU), and the Department
of Energy under Contract No. DE-FG02-04ER41291 (U. Hawaii).

\end{document}